\documentclass{LMCS}
\usepackage{amsmath}
\usepackage{amssymb}
\usepackage{latexsym}
\usepackage{graphicx}
\usepackage{epstopdf}
\usepackage{wasysym}
\usepackage[linesnumbered,boxed]{algorithm2e}
\usepackage{enumerate,hyperref}

\newcommand{\ie}{\textit{i.e.}\xspace}

\DeclareMathOperator{\M}{Max}
\DeclareMathOperator{\m}{Min}
\DeclareMathOperator{\att}{DetAtt}
\DeclareMathOperator{\val}{val}
\DeclareMathOperator{\reach}{Reach}
\newcommand{\game}{G}

\newcommand{\NN}{\mathbb{N}}
\newcommand{\vertices}{V}
\newcommand{\vM}{\vertices_{\M}}
\newcommand{\vm}{\vertices_{\m}}
\newcommand{\vr}{\vertices_{\textrm{R}}}
\newcommand{\vv}{{\mathbf{f}}}
\newcommand{\vvv}{{\mathbf{g}}}

\newcommand{\sM}{\sigma}
\newcommand{\sm}{\tau}
\newcommand{\sopt}{\sigma^\#}
\newcommand{\topt}{\tau^\#}
\newcommand{\edges}{E}
\newcommand{\tp}[2]{\delta(#2)(#1)}
\newcommand{\tpseul}{\delta}

\newcommand{\target}{\circledcirc}
\newcommand{\sink}{\otimes}
\newcommand{\detatt}[1]{\att(#1)}
\newcommand{\bh}{\setminus}
\newcommand{\sigmav}{\sigma_\vv}
\newcommand{\sigmavv}{\sigma_\vvv}

\newcommand{\tauv}{\tau_\vv}

\newcommand{\fplay}{h}

\newcommand{\proba}[4]{\mathbb{P}^{#1,#2}_{#3}\left(#4\right)}
\newcommand{\probaa}[3]{\mathbb{P}^{#1,#2}_{#3}}
\newcommand{\esper}[4]{\mathbb{E}^{#1,#2}_{#3}\left[#4\right]}

\newcommand{\mc}[1]{\mathcal{M}_{#1}}
\newcommand{\bigO}[1]{\operatorname{O}\bigl(#1\bigr)}
\newcommand{\complex}{\bigO{|\vr|!\cdot(|\edges|+|\tpseul|)}}

\renewcommand{\phi}{\varphi}

\SetKwFor{REPEAT}{repeat}{}{}
\SetVline

\def\doi{5 (2:9) 2009}
\lmcsheading%
{\doi}
{1--17}
{}
{}
{Jul.~10, 2008}
{May~\phantom{.}25, 2009}
{}   

\begin{document}

\title[Solving Simple Stochastic Games...]{Solving Simple Stochastic Games\\with Few Random Vertices}

\author[H.~Gimbert]{Hugo Gimbert\rsuper a}	%required
\address{{\lsuper a}LaBRI, CNRS, Bordeaux, France}	%required
\email{hugo.gimbert@labri.fr}  %optional

\author[F.~Horn]{Florian Horn\rsuper b}	%optional
\address{{\lsuper b}CWI, Amsterdam, The Netherlands}	%optional
\email{f.horn@cwi.nl}  %optional
\thanks{{\lsuper b}This research was partially supported by the french project ANR ``DOTS''. The second author held the tenure of an ERCIM ``Alain Bensoussan'' fellowship programme.}	%optional

%% etc.

%% required for running head on odd and even pages, use suitable
%% abbreviations in case of long titles and many authors:

%% mandatory lists of keywords and classifications:
%\titlecomment{}
%%%%%%%%%%%%%%%%%%%%%%%%%%%%%%%%%%%%%%%%%%%%%%%%%%%%%%%%%%%%%%%%%%%%%%%%%%%

%% the abstract has to PRECEED the command \maketitle:
%% be sure not to issue the \maketitle command twice!

\begin{abstract}
Simple stochastic games are two-player zero-sum stochastic games with turn-based moves, perfect information, and reachability winning conditions.

We present two new algorithms computing the values of simple stochastic games. Both of them rely on the existence of optimal \emph{permutation strategies}, a class of positional strategies derived from permutations of the random vertices. The ``permutation-enumeration'' algorithm performs an exhaustive search among these strategies, while the ``permutation-improvement'' algorithm is based on successive improvements, \emph{\`a la} Hoffman-Karp.

Our algorithms improve previously known algorithms in several aspects.
First they run in polynomial time when the number of random vertices is fixed, so the problem of solving simple stochastic games is fixed-parameter tractable when the parameter is the number of random vertices. Furthermore, our algorithms do not require the input game to be transformed into a stopping game. Finally, the permutation-enumeration algorithm does not use linear programming, while the permutation-improvement algorithm may run in polynomial time.
\end{abstract}

\maketitle

\section*{Introduction}

\begin{figure}[ht]

	\begin{center}
	\includegraphics{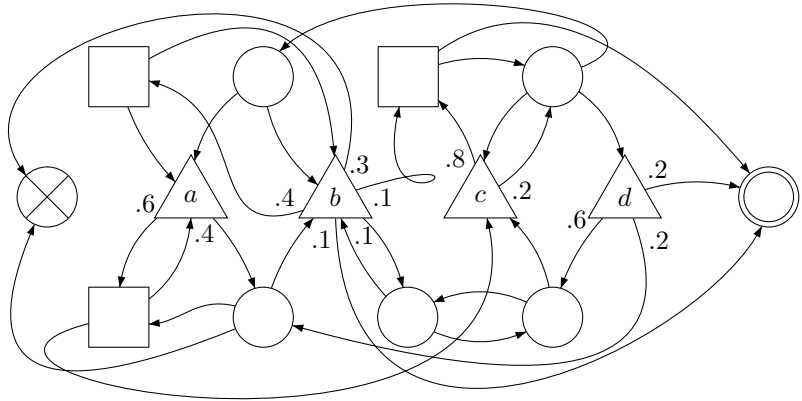}	
	\end{center}
	\caption{A Simple Stochastic Game.}
	\label{fig:exampleSSG}
\end{figure}

Simple stochastic games (SSGs) are played by two players called $\M$ and $\m$ in a sequence of steps. The players move a pebble along the edges of a directed graph $(\vertices,\edges)$ whose vertices are partionned into three sets: $\vM$, $\vm$, and $\vr$. When the pebble is on a vertex of $\vM$ or $\vm$, the corresponding player chooses an outgoing edge and moves the pebble along it. When the pebble is on a vertex of $\vr$ (a \emph{random} vertex), the outgoing edge is chosen randomly according to a fixed probability distribution. The players have opposite goals, as $\M$ wants to reach a special sink vertex $\target$ while $\m$ wants to avoid it forever. An example of SSG is depicted in Figure~\ref{fig:exampleSSG}, with vertices of $\vM$ represented as $\fullmoon$'s, vertices of $\vm$ represented as $\Box$'s, and vertices of $\vr$ represented as $\vartriangle$'s.

SSGs are a natural model of reactive systems. Consider, for example, a hardware component. It can be modelled as an SSG, whose vertices represent the global states of the component and the target is some error state to avoid. The nature of a given vertex depends on who can influence the immediate evolution of the system: it is a $\m$ vertex if the software can choose between different options, a $\M$ vertex if there is a (non-deterministic) input asked from the user, and a random vertex if the evolution depends on a stochastic environment. An optimal strategy for $\m$ can then be used as the basis for the synthesis of a ``good'' driver, \ie one which minimises the probability of entering the error state independently of the behaviour of the user.

%je ne vois jamais de ' dans les papiers ils sont implicites
%cf remarque plus bas aussi
The main algorithmic problem about SSGs is the computation of values of the vertices and optimal strategies for the players.
%It has been first studied by Condon,
%bof le  pronom personnel en debut de phrase.
This problem was first adressed by Condon,
 who showed that deciding whether the value of a vertex is greater than $\frac{1}{2}$ belongs to {\tt NP} and co-{\tt NP}~\cite{Con92-IC}.
 %Her algorithm
 %bof
 Condon's algorithm guesses non-deterministically the values of vertices,
%il faut choisir entre vertices' values ou values of vertices
which are rational numbers of linear size, and checks that they are solutions of some \emph{local optimality equations}.
%It supposes
%"it" bof: ca refere a quoi?
%+ un algorithme ne suppose rien
This algorithm is correct only for \emph{stopping} games,
where the pebble reaches either the target or a sink target with probability one, regardless of the players' strategies.
Any SSG can be transformed in polynomial time into a stopping SSG with (almost) the same values, but it incurs a quadratic blow-up
% in 
of the size of the game.

%Three other algorithms solving SSGs, presented in~\cite{Con93-DMTCS},
%also rely on the ``stopping'' hypothesis
% :  faux car le strategy-improvement et le quadratic algo
%sont corrects sans l'hypothèse de stopping property.
Three other algorithms for solving SSGs are presented in~\cite{Con93-DMTCS}.
 The first one computes the values of the vertices using a quadratic program with linear constraints. The second one computes iteratively from below the values of the vertices, and the third is a strategy improvement algorithm \emph{\`a la} Hoffman-Karp~\cite{HK66-MS}. The two latter algorithms, as the ones recently proposed in~\cite{Som05-ENTCS}, solve a series of linear programs which could be of exponential length. Furthermore, solving a linear program requires high-precision arithmetic, even if it can be done in polynomial time~\cite{Kha79-SMD,Ren88-MP}. The best randomised algorithms achieve sub-exponential expected time $e^{O(\sqrt{n})}$~\cite{Lud95-IC,Hal07-Alg}.
% Randomised algorithms do not perform much better: so far, the best randomised algorithms~\cite{Lud95-IC,Hal07-Alg} run in sub-exponential expected time $e^{O(\sqrt{n})}$.

In this paper we present two algorithms computing the values and optimal strategies in SSGs: the ``permutation-enumeration'' and the ``permutation-improvement'' algorithms. The common basis for both algorithms is that optimal strategies can be looked for in a subset of the positional strategies
%,
called \emph{permutation strategies}.
% These:
%refere a "both" algorithms ou a "positional strategies" ?
Permutation strategies are derived from permutations over the random vertices. In order to find optimal strategies, the permutation-enumeration algorithm performs an exhaustive search among all permutation strategies, whereas the permutation-improvement algorithm performs successive improvements of permutation strategies, \emph{\`a la} Hoffman-Karp~\cite{HK66-MS}.

%They
%tu ne peux pas mettre un pronom en debut de paragraphe
%ca empeche la lecture en diagonal de l'intro
%redondances dans contenu et style = style lourd et papier lisible et precis
The permutation-enumeration and the permutation-improvement algorithms share two advantages over existing algorithms.
First, they perform much better on SSGs with few random vertices, as they run in polynomial time when the number of random vertices is logarithmic in the size of the game: it follows that the problem of solving SSGs is fixed-parameter tractable when the parameter is the number of random vertices.
Second, they do not rely on the
%expensive and unintuitive
%c'est très bien de critiquer le travail des autres pour mettre ce qu'on fait en perspective
%par contre le faire de maniere subjective sans argument precis ne donne pas d'info au lecteur
transformation of the input SSG into a stopping SSG,
which avoids the quadratic blow-up of the size of the game.
%c'est ça l'argument precis
%j'avais rajoute comme argument que ca permettait d'adapter l'algo a d'autres jeux avec ref
%a ta these mais tu as supprime ça dommage
Moreover, the permutation-enumeration algorithm does not use linear or quadratic programming,
%(it just computes the values of Markov chains)
%les gens, (nous deux avant l'ecriture de cet article par exemple)
%ne savent pas comment resoudre une chaine de Markov
%ils pensent que c'est complique parce qu'il y a "Markov" dans le nom
(it just computes the solutions to linear systems)
and its worst-case complexity is $\complex$, where $|\vr|$ is the number of random vertices, $|\edges|$ is the number of edges and $|\tpseul|$ is the maximal bit-length of transition probabilities. The nominal complexity of the permutation-improvement algorithm is higher
%"nominal complexity" d'un algorithme, c'est quoi?
but we
%don't
%pas de "don't" a l'ecrit
do not know any non-trivial lower bound for
%the 
its complexity:
%of the permutation-improvement algorithm
%it
the permutation-improvement algorithm may actually run in polynomial time.

\medskip

\noindent \textbf{Outline.} In Section~\ref{sec:defs}, we provide formal definitions for SSGs, values and optimal strategies. We describe then in Section~\ref{sec:playing} the central notion of permutation strategies. Section~\ref{sec:optimality} presents the permutation-enumeration algorithm, based on the \emph{self-consistency} and \emph{liveness} properties. Section~\ref{sec:heuristics} introduces an improvement policy for permutations which leads to the permutation-improvement algorithm.

\section{Simple Stochastic Games}
\label{sec:defs}

% In this section we recall the main notions and results about simple stochastic games.

\subsection{Plays and strategies}

A \emph{simple stochastic game} is a tuple $(\vertices, \vM, \vm, \vr,E,\tpseul,\target)$, where $(\vertices,\edges)$ is a graph, $(\vM,\vm,\vr)$ is a partition of $\vertices$, and $\target$ is a distinguished sink vertex in $\vertices$ called the \emph{target} of the game. The transitions from the random vertices are equipped with probabilities described by the function $\tpseul : \vr \rightarrow \vertices \rightarrow [0,1]$, such that for all $v\in\vr$, $w \in \vertices$, $\tp{w}{v} > 0 \Rightarrow (v,w) \in \edges$, and $\sum_{w\in\vertices} \tp{w}{v} =1$.

An \emph{infinite play} $\rho$ is an infinite sequence $\rho_0\rho_1\dots\in\vertices^\omega$ of vertices such that for all $i \in \mathbb{N}, (\rho_i,\rho_{i+1})\in\edges$. It is \emph{winning for $\M$} if there is a $i \in \NN$ such that $\rho_i = \target$ (as $\target$ is a sink, it follows that $\forall j > i, \rho_j = \target$). Otherwise, $\rho$ is \emph{winning for $\m$}. A \emph{finite play} is a finite prefix of an infinite play.

A (pure) \emph{strategy} for $\M$ is a mapping $\sM:\vertices^*\vM \to \vertices$ such that for each finite play $\fplay = \fplay_0 \ldots \fplay_i$ ending in a $\M$ vertex, $(\fplay_i,\sigma(\fplay))\in\edges$. It is \emph{positional} if it only depends on the last vertex of $h$: $\sigma(h) = \sigma(h_i)$. A play $\rho_0\rho_1\ldots$ is \emph{consistent with $\sM$} if for every $i$ such that $\rho_i\in\vM$, $\rho_{i+1} = \sigma(\rho_0 \dots \rho_i)$. Strategies for $\m$ are defined analogously and are generally denoted by $\sm$.

\subsection{Measures and values}

The set of plays is made into a measurable space on the $\sigma$-algebra generated by the canonical projections $\{V_i\}_{i\in\NN}$, where $V_i(\rho_0\rho_1\dots)=\rho_i$~\cite{Bil95}. Once an initial vertex $v$ and two strategies $\sM$ and $\sm$ for players $\M$ and $\m$ have been fixed, the probability measure $\probaa{\sigma}{\tau}{v}$ is defined by:
\begin{align*}
&\proba{\sigma}{\tau}{v}{V_0=v}=1\enspace,\\
&\proba{\sigma}{\tau}{v}{V_{i+1} = \sigma(V_0\dots V_i)\mid V_i\in\vM}=1\enspace,\\
&\proba{\sigma}{\tau}{v}{V_{i+1} = \tau(V_0\dots V_i)\mid V_i\in\vm}=1\enspace,\\
&\proba{\sigma}{\tau}{v}{V_{i+1}\mid V_i \in \vr} = \tp{V_{i+1}}{V_i}\enspace.
\end{align*}
The expectation of a real-valued, measurable and bounded function $\phi$ under $\probaa{\sigma}{\tau}{v}$ is denoted $\esper{\sigma}{\tau}{v}{\phi}$. We will often use implicitly the following formulae which rule the probabilities and expectations once a finite prefix $h=h_0 \ldots h_i$ is fixed:
\begin{eqnarray}
\probaa{\sigma}{\tau}{v}(\Gamma \mid V_0\dots V_i = h_0 \dots h_i) & = &\probaa{\sigma[h]}{\tau[h]}{h_i}(\Gamma[h])\enspace, \label{eq:pdecal}\\
\esper{\sigma}{\tau}{v}{\phi\mid V_0\dots V_i = h_0 \dots h_i} & = &\esper{\sigma[h]}{\tau[h]}{h_i}{\phi[h]}\enspace, \label{eq:edecal}
\end{eqnarray}
%where $\sigma[h](\rho_0\rho_1\ldots) = \sigma(h_0\ldots h_i \rho_0\rho_1\ldots)$, and $\tau[h]$, $\Gamma[h]$, and $\phi[h]$ are defined analogously.
%ma correction etait bonne le dernier etat de h et le premier de rho coincident
where $\sigma[h](\rho_0\rho_1\ldots) = \sigma(h_0\ldots h_{i-1} \rho_0\rho_1\ldots)$, and $\tau[h]$, $\Gamma[h]$, and $\phi[h]$ are defined analogously.

If we fix only $\M$'s strategy $\sigma$ and the initial vertex $v$, the target vertex will be reached with probability at least:
	\begin{displaymath}
		\inf_{\tau} \proba{\sigma}{\tau}{v}{\reach(\target)}\enspace,
	\end{displaymath}
where $\reach(\target)$ is the event $\{ \exists i\in\NN, V_i = \target\}$.
%Starting from $v$, $\M$ has strategies which ensure a probability to win arbitrarily close to:
%faux car min peut s'amuser a se sacrifier
%"ensure a probability" pas de l'anglais
Starting from $v$, player $\M$
has strategies that guarantee a winning outcome with a probability greater than:
	\begin{displaymath}
		\val_*(v) = \sup_{\sigma} \inf_{\tau} \proba{\sigma}{\tau}{v}{\reach(\target)}\enspace,
	\end{displaymath}
minus $\epsilon$ for any $\epsilon>0$.
Symmetrically, $\m$ has strategies
%to bound the probability that $\M$ wins arbitrarily close to:
%je ne comprends pas
that guarantee a winning outcome with a probability less than:
	\begin{displaymath}
		\val^*(v) =  \inf_{\tau}\sup_{\sigma} \proba{\sigma}{\tau}{v}{\reach(\target)}\enspace,
	\end{displaymath}
plus $\epsilon$ for any $\epsilon>0$.
It is clear that $\val_*(v) \leq \val^*(v)$. In the case of SSGs, stronger results are known:
\begin{thm}[\cite{Sha53-PNAS,Gil57,LL69-SIAM}]
% ,Gil57,LL69-SIAM}]
\label{theo:pos}
Let $\game = (\vertices, \vM, \vm, \vr,E,\target,\tpseul)$ be a SSG. Then, for any vertex $v \in \vertices$, 
\begin{displaymath}
	\val_*(v) = \val^*(v) \enspace .
\end{displaymath}
This common value is denoted by $\val(v)$. Furthermore, there are positional optimal strategies for both players, \ie positional strategies $\sopt$ and $\topt$ such that, for any strategies $\sigma$ and $\tau$:
\begin{displaymath}
	\proba{\sigma}{\topt}{v}{\reach(\target)} \leq \val(v) \leq \proba{\sopt}{\tau}{v}{\reach(\target)}\enspace .
\end{displaymath}
\end{thm}

\subsection{Normalised games} 
\begin{figure}[ht]
%\unitlength=1.5mm
\begin{center}
\includegraphics{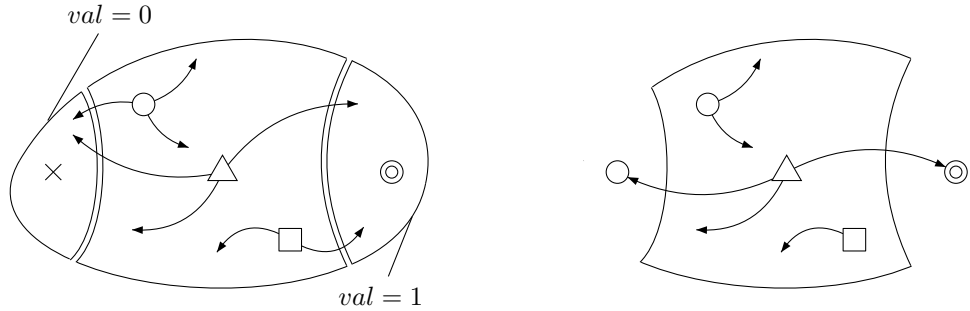}
\end{center}
\caption{\label{fig:normalisation}Normalisation.}
\end{figure}

A SSG is \emph{normalised} if the only vertex with value 1 is the target $\target$ and there is only one (sink) vertex $\sink$ with value $0$. Our motivations for the introduction of this notion are twofold. First, several proofs are much simpler for normalised games.
Second, any SSG can be reduced to an equivalent normalised game in linear time
%, 
and the resulting game is smaller than the original one.
This reduction is presented on Figure~\ref{fig:normalisation}: it simply consists in merging the region with value one into $\target$ and the region with value zero into a new sink vertex $\sink$.

In the remainder of this article, we assume that we are working on a normalised SSG $G =(\vertices, \vM, \vm, \vr,E,\tpseul,\target,\sink)$, with $k$ random vertices.

\section{Permutation strategies}
\label{sec:playing}
	
%The existence of positional optimal strategies is a key property of SSGs and the cornerstone of many algorithms solving them.
%pas clair a quoi le them referre
The existence of positional optimal strategies is a key property of SSGs and the cornerstone of many algorithms solving these games.
The algorithms we propose rely on a refinement of this result: optimal strategies can be looked for among a subset of the positional strategies,
the set of ``permutation strategies''.

As a matter of fact, Theorem~\ref{theo:pos} is a corollary of results of the present paper.
The proofs of our results often rely on the existence of values and optimal (not only $\epsilon$-optimal) strategies in SSGs.
This could be avoided ---the main point is to use $\val_*$ instead of $\val$---
but we felt that it was not worth the extra complexity.
	
\begin{figure}[ht]
\begin{center}
\includegraphics{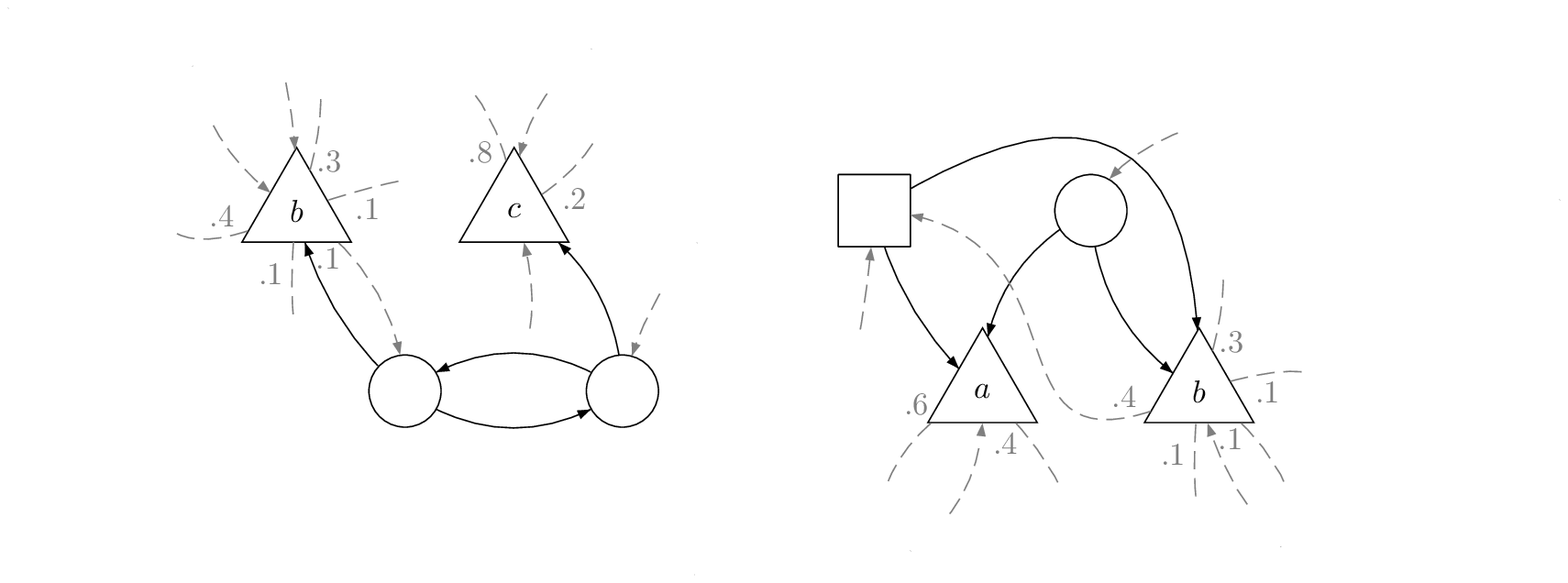}
\end{center}
\caption{\label{fig:intuitions}Coherence and contention.}
\end{figure}

The main intuition underlying permutation strategies is that the only meaningful events in a play are the visits to random vertices. Between two visits the players only strive to impose which random vertex will be visited next, and the result of their interaction can be easily predicted. This is illustrated by Figure~\ref{fig:intuitions}, which zooms on two details of Figure~\ref{fig:exampleSSG}.
	
	In the left part of Figure~\ref{fig:intuitions}, $\M$ must choose between the two random vertices $b$ and $c$ (refusing to choose is not really an option). There is no reason to choose $b$ in one of the vertices, and $c$ in the other. We could consider only the strategies ``always go to $b$'' and ``always go to $c$''. In the right part of Figure~\ref{fig:intuitions}, we consider relationships between the two players' strategies. From their respective vertices $\fullmoon$ and $\Box$, $\M$ and $\m$ can send the pebble to either $a$ or $b$. We could restrict our attention to the cases where $\M$ goes to one, and $\m$ to the other.

	Underlying these intuitions is the idea of a ``preference order'' over the random vertices. In the remainder of this article, we formalise it as a \emph{permutation}: a 
%bijection
%est peu utilisé en anglais
one-to-one correspondance $\vv$ between $\vr$ and $\{1,\ldots,k\}$. For simplicity, we often write $\vv_i$ instead of $\vv^{-1}(i)$
and we consider the sink and target vertices as random vertices
with the implicit assumption that they are respectively the lowest and greatest vertices: $\vv_0 = \sink$ and $\vv_{k+1} = \target$.
	
	\subsection{Attractors and $\vv$-regions}

		Once a permutation $\vv:\vr\to\{1,\ldots,k\}$ has been fixed, the $\vv$-strategies consist in $\M$ trying to reach the highest (with respect to $\vv$) possible random vertex, while $\m$ tries to thwart her. Notice that the situation is not exactly symmetric, since the burden of reaching a random vertex lies with $\M$:
%$\m$ wins if
% suggere un iff mais c'est pas le cas
in case the pebble remains forever in controlled vertices then player $\m$ wins.
The formal definition of permutation strategies is based on the notion of \emph{deterministic attractor}.

		\begin{defi}
		\label{defi:detatt}
			Let $X \subseteq \vertices$ be a set of vertices. The deterministic attractor of $\M$ to $X$, denoted by $\detatt{X}$, is computed recursively:
				\begin{eqnarray*}
				X^0 & = & X \enspace ,\\
				X^{i+1} & = & X^i \cup \left\{v \in \vM \mid \exists w \in X^i, (v,w) \in \edges\right\}\\
					&   & \hphantom{X^i} \cup \left\{v \in \vm \mid \forall w \in \vertices, (v,w) \in \edges \Rightarrow w \in X^i\right\} \enspace ,\\
				\detatt{X} & = & \bigcup_{i>0} X^i \enspace .
				\end{eqnarray*}
			An attracting strategy to $X$ for $\M$ is a positional strategy $\sigma$ such that:
				\begin{displaymath}
					\forall i \ge 1, \sigma(X^i) \subseteq X^{i-1} \enspace .
				\end{displaymath}
			Symmetrically, a trapping strategy out of $X$ for $\m$ is a positional strategy $\tau$ such that:
				\begin{displaymath}
					\tau(\vertices \setminus \detatt{X}) \subseteq \vertices \setminus \detatt{X} \enspace .
				\end{displaymath}
		\end{defi}

		The \emph{$\vv$-regions} associated with a permutation $\vv:\vr\to\{1,\ldots,k\}$
		are defined as embedded deterministic attractors to the random vertices: 
			\begin{eqnarray*}
				W_\vv[k+1] & = & \{\target\} \enspace ,\\
				\forall 1 \le i \le k, W_\vv[i] & = & \detatt{\{\vv_i,\ldots, \vv_k,\target\}} \setminus \bigcup_{j>i} W_\vv[j]\enspace ,\\
				W_\vv[0] & = & \{\sink\} \enspace .
			\end{eqnarray*}

%We have then the following properties:
%"We have then" bof
%on en sait pas encore ce que sont les vv strats
\subsection{Permutation strategies}
The \emph{$\vv$-strategies} $\sigmav$ and $\tauv$ are strategies such that, on each $W_\vv[i]$:
	\begin{enumerate}[$\bullet$]
		\item	$\sigmav$ coincides with an attractor strategy to $\{\vv_i,\ldots, \vv_k,\target\}$,
		\item	$\tauv$ coincides with a trapping strategy out of $\{\vv_{i+1},\ldots, \vv_k,\target\}$.
	\end{enumerate}

\begin{figure}[ht]

	\begin{center}
	\includegraphics[scale=.95]{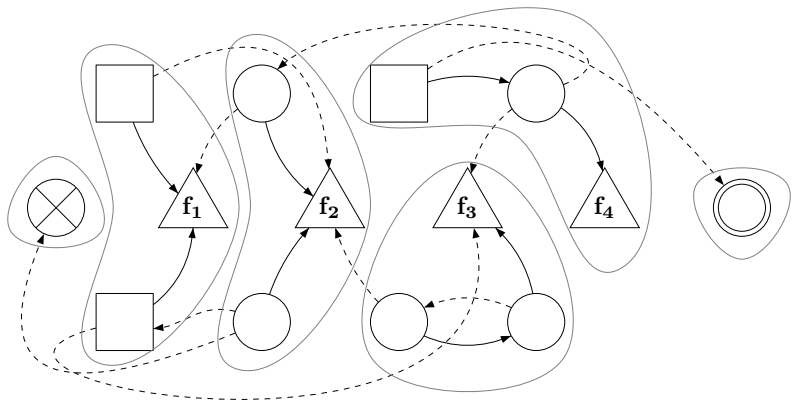}

	\end{center}
	\caption{$\vv$-regions and $\vv$-strategies in the game of Figure~\ref{fig:exampleSSG}.}
	\label{fig:regions}
\end{figure}

		The $\vv$-regions partition $\vertices$, so we extend the definition domain of $\vv:\vr\to\{1,\ldots,k\}$ to $\vertices$ in a natural way:
$\vv(v) = i$ if $v \in W_\vv[i]$.
The following properties are easy to prove:
		\begin{align}
			&\forall v \in \vM, && \vv(v) = \vv(\sigmav(v)) \enspace ,\label{eq:ext1}\\
			&\forall v \in \vM, \forall (v,w) \in \edges, && \vv(v) \ge \vv(w) \enspace ,\label{eq:ext2}\\
			&\forall v \in \vm, && \vv(v) = \vv(\tauv(v)) \enspace ,\label{eq:ext3}\\
			&\forall v \in \vm, \forall (v,w) \in \edges, && \vv(v) \le \vv(w) \enspace .\label{eq:ext4}
		\end{align}
If $\M$ plays $\sigmav$ and $\m$ plays $\tauv$ from an initial vertex $v$, the first random vertex reached by the pebble is the unique random vertex
$w$ such that $\vv(w) = \vv(v)$.
Figure~\ref{fig:regions} describes the $\vv$-regions and $\vv$-strategies of the game of Figure~\ref{fig:exampleSSG}, for $\vv = abcd$.

	\subsection{The $\vv$-values}

When both players use their respective permutation strategies, the probability that a pebble starting in $v$ reaches $\target$ is denoted by $\phi_\vv(v)$:
		\begin{displaymath}
			\phi_\vv(v) = \probaa{\sigmav}{\tauv}{v}(\reach(\target)) \enspace .
		\end{displaymath}

	\begin{prop}
	\label{prop:calcstrat}
		Let $\vv$ be a permutation. The $\vv$-regions and the $\vv$-strategies can be computed in time $\bigO{|\edges|\log^*(|\vertices|)}$ and
		the $\vv$-values can be computed in time $\bigO{|\vr|^3|\tpseul|}$.
	\end{prop}

	\begin{proof}
	The $\vv$-regions and $\vv$-strategies can be expressed in terms of
%\emph{non-stochastic} games,
%n'existent pas. J'ai deja fait cette erreur
\emph{deterministic} games as they do not depend on what happens once a random vertex is reached. We can thus use the results of~\cite{AHMS08-CIE} to compute them in time $\mathcal{O}(|E|\log^*(|\vertices |)$. In order to compute the $\vv$-values, we build a Markov chain $\mc{\vv}$ designed to mimic the behaviour of $G$ when the players use their $\vv$-strategies. Intuitively, we merge each region $W_\vv[i]$ into a single vertex $i$; formally, $\mc{\vv}$ is a Markov chain with states $S = \{0,\ldots,k+1\}$ such that $0$ and $k+1$ are absorbing and, for every $1 \le i \le k$ and $0 \le j \le k+1$, the transition probability from $i$ to $j$ is given by:
		\begin{displaymath}
			p_{ij} = \sum_{v\in W_\vv[j]}\tp{v}{\vv_i} \enspace .
		\end{displaymath}

	The values $x^* : \{0,\ldots,k+1\} \rightarrow [0,1]$ of $\mc{\vv}$ are computed as follows. Let $I\subseteq S$ be the set of vertices from which $k+1$ is reachable in $\mc{\vv}$. Then, for each $i \notin I$, $x^*_i = 0$, and $(x^*_{i})_{i\in I}$ is the unique solution of the following linear system:
	\begin{equation*}
		\begin{cases}
		x^*_{k+1}&=1\\
		x^*_{i} &= \sum_{j\in I} p_{i,j} \cdot x^*_{j}\enspace,
		\end{cases}
	\end{equation*}
	which can be solved in time $\bigO{|\vr|^3|\tpseul|}$~\cite{Dix82-NM}. For each $v \in \vertices$, $\phi_\vv(v) = x^*_{\vv(v)}$.
\end{proof}

\section{The permutation-enumeration algorithm}
\label{sec:optimality}

%The main point about permutation strategies is the following theorem:
%c'est nul comme debut de section il faut dire au lecteur ce qui l'attend

In this section we describe the permutation-enumeration algorithm which computes optimal strategies for both players.
This algorithm relies on the following key property of permutation strategies.

\begin{thm}
\label{thm:vstrat1}
In every SSG, there exists a permutation $\vv$
%over the random vertices
such that $\sigmav$ is optimal for $\M$ and $\tauv$ is optimal for $\m$.
\end{thm}

%It
%What is it?
%
%suggests a very simple enumerative algorithm checking for each permutation $\vv$ whether the $\vv$-strategies are optimal: each test can be performed %in polynomial time using linear programming~\cite{Der72,Con92-IC}.
%Tu oublies de dire a quoi sert cet algorithm, le lecteur est oblige de deviner, et ce en debut de section en plus (lecture diagonale...)
This theorem suggests a very simple enumerative algorithm computing values and optimal strategies:
check for each permutation $\vv$ whether the $\vv$-strategies are optimal.
Each test can be performed in polynomial time using linear programming~\cite{Der72,Con92-IC}.
However, linear programming requires high-precision arithmetic and is expensive in practice.
Our permutation-enumeration algorithm uses a simpler criterion based on a refinement of Theorem~\ref{thm:vstrat1}:
we look for permutations which are \emph{live} and \emph{self-consistent}.
%tres bien

\subsection{Liveness and self-consistency}
\label{sec:testing}

The permutation-enumeration algorithm is based on two simple properties on permutations: self-consistency and liveness.
Self-consistency expresses the adequation between \emph{a priori} preferences (permutation $\vv$)
and resulting values (the $\vv$-values $\phi_\vv$).
Liveness stipulates that each random vertex has a positive probability to immediately lead to a better ---from $\M$'s point of view--- region.

\begin{defi}
\label{defi:selfconsistency}
	A permutation $\vv$ is self-consistent if:
		\begin{displaymath}
		\phi_\vv(\vv_1) \le \phi_\vv(\vv_2) \le \ldots \le \phi_\vv(\vv_k) \enspace.
		\end{displaymath}
\end{defi}

\begin{defi}
\label{defi:liveness}
	A permutation $\vv$ is live if:
	\begin{displaymath}
	 	\forall 1 \le i \le k, \exists j > i, \exists v \in W_\vv[j], \tp{v}{\vv_i} > 0\enspace.
	\end{displaymath}
\end{defi}

As we show below, the $\vv$-strategies associated with a live and self-consistent permutation $\vv$ are optimal and there is always such a permutation. The permutation-enumeration algorithm
%is then
%pourquoi "then"?
performs an exhaustive search for a live and self-consistent permutation.

\begin{algorithm}[ht]
\KwIn{A normalised simple stochastic game $G=(\vertices,\vM,\vm,\vr,\edges,\tpseul,\target,\sink)$.}
\KwOut{Optimal strategies for $\M$ and $\m$.}
	\ForAll{permutation $\vv$ over $\vr$}{
		compute the $\vv$-regions\;
%		compute the values of the Markov chain $\mc{\vv}$\;
%mc vv n'apparait pas dans l'enonce de la prop 2.2
		compute the $\vv$-values\;
		\If{$\vv$ is live and self-consistent}{
			 \Return{$\sigmav$ and $\tauv$}\;
		}
}
\caption{The permutation-enumeration algorithm.}
\end{algorithm}

\begin{thm}\label{theo:algo2}
The permutation-enumeration algorithm terminates and returns optimal strategies for $\M$ and $\m$. Its worst-case running time is $\complex$.
\end{thm}

\begin{proof}
	Correctness and termination are proved in Lemmas~\ref{lem:correctness} and~\ref{lem:existence}, respectively. The worst-case complexity follows from the fact that there are at most $k!$ permutations and Proposition~\ref{prop:calcstrat}.
\end{proof}

\begin{figure}[ht]

	\begin{center}
		\includegraphics{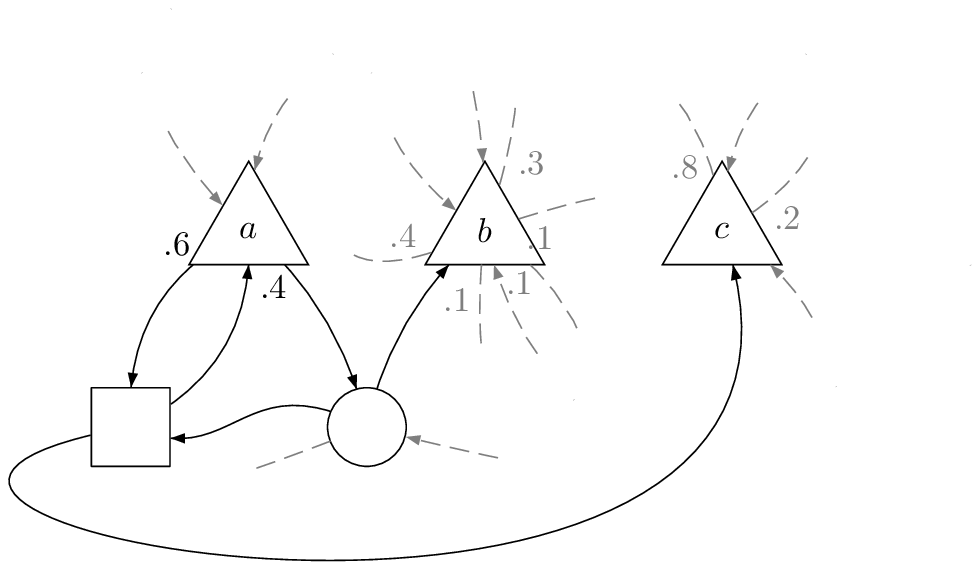}
	\end{center}
	\caption{Self-consistency does not guarantee optimality.}
	\label{fig:caseliveness}
\end{figure}

Before we proceed with the proofs of the main lemmas, let us make a case for liveness: Figure~\ref{fig:caseliveness} shows that self-consistency is not enough to guarantee the optimality of the resulting strategies\footnote{It \emph{would} be enough in stopping games, but testing liveness is cheaper than the reduction.}.

In this excerpt from the game of Figure~\ref{fig:exampleSSG}, $\M$'s strategy in $\fullmoon$ should be to send the pebble to $b$, as $\m$ could otherwise trap the play in $\{a,\fullmoon,\Box\}$. However, consider the permutation $\vvv = bcad$: $\m$ sends the pebble from $\Box$ to $c$ to avoid $a$; $\M$ sends the pebble from $\fullmoon$ to $\Box$ to reach either $a$ or $c$. We have thus $\phi_\vvv(a) = \phi_\vvv(c)$. As a matter of fact, we have $\phi_\vvv(b) \leq \phi_\vvv(a) = \phi_\vvv(c) \le \phi_\vvv(d)$, so $\vvv$ is self-consistent even though the $\vvv$-values are not the correct ones. Liveness forbids this kind of gambits from $\M$. It replaces, in this aspect, the ``stopping'' hypothesis of Condon.

\subsection{Correctness of the permutation-enumeration algorithm} We first show that if a permutation $\vv$ is live and self-consistent, the $\vv$-strategies are optimal (Lemma~\ref{lem:correctness}). We need two preliminary propositions. First, if $\vv$ is self-consistent and $\M$ plays according to $\sigmav$, the sequence $(\phi_\vv(V_i))_{i\in\NN}$ is a submartingale\footnote{We do not use any result about martingales in this paper.} and symmetrically if $\vv$ is self-consistent and $\m$ plays according to $\tauv$
%,
the sequence $(\phi_\vv(V_i))_{i\in\NN}$ is a supermartingale.

\begin{prop}\label{prop:martingale}
Let $\vv$ be a self-consistent permutation. Then, for any strategies $\sigma$ and $\tau$ for $\M$ and $\m$, vertex $v$, and integer $i$,
\begin{align}
\label{eq:submartingale}
&\esper{\sigmav}{\tau}{v}{\phi_\vv(V_{i+1}) \mid V_0\dots V_i} \geq \phi_\vv(V_i)\enspace,\\
\label{eq:supermartingale}
&\esper{\sigma}{\tauv}{v}{\phi_\vv(V_{i+1}) \mid V_0\dots V_i} \leq \phi_\vv(V_i)\enspace.
\end{align}
\end{prop}

\begin{proof}
In order to prove ($\ref{eq:submartingale}$), it is enough to show that for any finite play $h = h_0 \ldots h_i$,
	\begin{eqnarray}
	\esper{\sigmav}{\tau}{v}{~\phi_\vv(V_{i+1})~\mid V_0\dots V_i=h_0\dots h_i} & \geq & \phi_\vv(h_i)\enspace. \label{eq:sub1}
	\end{eqnarray}
Depending on the owner of $h_i$, (\ref{eq:sub1}) follows from one of the following properties of $\phi_\vv$:
\begin{align}
&\forall v \in \vM, &&\phi_\vv(v) = \phi_\vv(\sigmav(v)) \label{eq:ttt2} \enspace,\\
&\forall v \in \vm, \forall (v,w)\in\edges, &&\phi_\vv(v) \leq \phi_\vv(w) \label{eq:ttt5} \enspace,\\
&\forall v \in \vr, &&\phi_\vv(v) = \sum_{w\in\vertices} \tp{w}{v}\cdot\phi_\vv(w) \label{eq:ttt1}\enspace.
\end{align}
The equations (\ref{eq:ttt2}) and (\ref{eq:ttt1}) follows from the definition of $\phi_\vv$, and (\ref{eq:ttt5}) follows from the self-consistency of $\vv$: by definition of the $\vv$-regions, if $v \in \vm$ and $(v,w) \in \edges$
%,
then $\vv(v) \le \vv(w)$ (see~\eqref{eq:ext2}), so $\phi_\vv(v) \le \phi_\vv(w)$. The proof of~\eqref{eq:supermartingale} is similar and we do not detail it.
\end{proof}

Second, we show a ``stopping property'' in the case where $\vv$ is live and $\M$ plays $\sigmav$.

\begin{prop}\label{prop:live}
Let $\vv$ be a live permutation. Then, for any strategy $\tau$ for $\m$ and initial vertex $v$,
	\begin{equation*}
		\proba{\sigmav}{\tau}{v}{\reach(\target) \vee \reach(\sink)} = 1\enspace.
	\end{equation*}
\end{prop}

\begin{proof}
	By definition of liveness, 
	\begin{displaymath}
%		\alpha = \min_{1 \le i \le k}\{\tp{\bigcup_{j>i}W_\vv[j]}{\vv_i}\}
%notation pas introduite
		\alpha = \min_{1 \le i \le k} \sum_{w\in\bigcup_{j>i}W_\vv[j]} \tp{w}{\vv_i}	\end{displaymath}
	is positive.
%If we denote $|\vertices|$ by $n$,
Let $n= |\vertices|$ and $k=|\vr|$ then
the definition of $\sigmav$ yields:
	\begin{equation*}
		\probaa{\sigmav}{\tau}{v}(V_n = \target \mid V_0 \neq \sink) \ge \alpha^k \enspace, \label{eq:live1}
	\end{equation*}
	or, since $\target$ and $\sink$ are sinks:
	\begin{equation*}
		\probaa{\sigmav}{\tau}{v}(\forall m\leq n, V_m \notin \{\target,\sink\}) \le 1-\alpha^k \enspace.
	\end{equation*}
	Equation (\ref{eq:pdecal}) yields:
	\begin{equation*}
		\forall i \in \NN, \probaa{\sigmav}{\tau}{v}(\forall m\leq {i \cdot n}, V_m \notin \{\target,\sink\}) \le (1-\alpha^k)^i \enspace,
	\end{equation*}
	hence $\probaa{\sigmav}{\tau}{v}(\forall m\in\NN, V_m \notin \{\target,\sink\})=0$ hence Proposition~\ref{prop:live}.
	%vraiment pas besoin de Borel-Cantelli
	% la sigma-continuite est un des axiomes definissant une mesure de proba
	%qui asssure la convergence 
\end{proof}

We can now prove the correctness of the permutation-enumeration algorithm:

\begin{lem}
\label{lem:correctness}
	Let $\vv$ be a live and self-consistent permutation. Then $\sigmav$ is optimal for $\M$ and $\tauv$ is optimal for $\m$.
\end{lem}

\begin{proof}
We first prove that $\sigmav$ ensures that a pebble starting from $v$ has probability at least $\phi_\vv(v)$ to reach $\target$:
\begin{eqnarray}
\proba{\sigmav}{\tau}{v}{\reach(\target)}	& = &	\esper{\sigmav}{\tau}{v}{\lim_{i\in\NN} \phi_\vv(V_i)}\label{eq:co2}\\
					& = &	\lim_{i\in\NN} \esper{\sigmav}{\tau}{v}{\phi_\vv(V_i)}\label{eq:co3}\\
					& \ge & \esper{\sigmav}{\tau}{v}{\phi_\vv(V_0)} = \phi_\vv(v)\label{eq:co4}\enspace,
\end{eqnarray}
where (\ref{eq:co2}) comes from Proposition~\ref{prop:live}, (\ref{eq:co3}) is a property of expectations, and (\ref{eq:co4}) comes from Proposition~\ref{prop:martingale}.

Then, we show that $\tauv$ ensures that a pebble starting from $v$ has probability at most $\phi_\vv(v)$ to reach $\target$:
\begin{eqnarray}
\proba{\sigma}{\tauv}{v}{\reach(\target)}	& \le &	\esper{\sigma}{\tauv}{v}{\liminf_{i\in\NN} \phi_\vv(V_i)}\label{eq:co5}\\
					& \le &	\liminf_{i\in\NN} \esper{\sigma}{\tauv}{v}{\phi_\vv(V_i)}\label{eq:co6}\\
					& \le &	\esper{\sigma}{\tauv}{v}{\phi_\vv(V_0)} = \phi_\vv(v)\label{eq:co7}\enspace,
\end{eqnarray}
where (\ref{eq:co5}) holds because $\target$ is a sink and $\phi_\vv(\target) = 1$, (\ref{eq:co6}) is a property of expectations, and (\ref{eq:co7}) comes from Proposition~\ref{prop:martingale}.

Thus, for any strategies $\sigma$ and $\tau$ for $\M$ and $\m$, 
\begin{displaymath}
	\proba{\sigma}{\tauv}{v}{\reach(\target)} \le \proba{\sigmav}{\tauv}{v}{\reach(\target)} \le \proba{\sigmav}{\tau}{v}{\reach(\target)} \enspace ,
\end{displaymath}
which completes the proof of Lemma~\ref{lem:correctness}
\end{proof}

\subsection{Termination of the permutation-enumeration algorithm}

Now we show the existence of a live and self-consistent permutation (Lemma~\ref{lem:existence}). Our construction is based on Proposition~\ref{prop:rec} and its correctness on Proposition~\ref{prop:valphi}.

\begin{prop}\label{prop:rec}
Let $X \subseteq \vertices$ be a set of vertices including the target vertex $\target$ and $Y$ be $\vertices \setminus \detatt{X}$. Then either $Y = \{\sink\}$ or there is a random vertex $v$ in $Y$ such that:
\begin{align*}
&\val(v)=\max \{\val(w)\mid w \in  Y\} \enspace ,\\
&\exists w \in \detatt{X}, \tp{w}{v} > 0 \enspace .
\end{align*}
\end{prop}

\begin{proof}
	Let $Z$ be the set of vertices with maximal value in $Y$:
	\begin{displaymath}
		Z = \{ v \in Y \mid \val(v) = \max_{w\in Y} \val(w)\} \enspace ,
	\end{displaymath}
	and suppose that:
	\begin{displaymath}
		\forall v \in \vr \cap Z, \forall w \in \detatt{X}, \tp{w}{v}=0 \enspace .
	\end{displaymath}
	Let $v$ be a vertex in $Z$.
	As $\game$ is normalised, we just need to show that $\val(v) = 0$, \ie there is a strategy $\theta$ for $\m$ such that
	for every strategy $\sigma$ for $\M$, $\probaa{\sigma}{\theta}{v}(\reach(\target)) = 0$.
	% be a positional optimal strategy 
	%inutile

	By definition of $\detatt{X}$, there is a positional strategy $\tau$ for $\m$ such that $\tau(Y) \subseteq Y$, and it follows from the definition of $Z$ that $\tau(Z) \subseteq Z$. As $Z$ is also closed under random moves, a pebble starting in $Z$ can only leave $Z$ through a move of $\M$, which leads to $Y\setminus Z$ as $Y = \vertices \setminus \detatt{X}$.

	We define now a non-positional strategy $\theta$ in which $\m$ plays according to $\tau$ as long as the play remains in $Z$ and switches definitively to an
	%positional
	%inutile
optimal strategy the first time the pebble moves out of $Z$. We can thus partition the plays starting in $v$ and consistent with $\sigma$ and $\theta$ depending on if and where the play gets out of $Z$: $\Gamma_Z$ is the set of plays remaining forever in $Z$, and for each $w$ in $Y \setminus Z$, $\Gamma_w$ is the set of plays where $w$ is the first visited vertex outside of $Z$.  Clearly %	\begin{eqnarray*}
$\probaa{\sigma}{\theta}{v}(\reach(\target) \mid \Gamma_Z) =0$
and by definition of the strategy $\theta$, $\forall w\in Y\bh Z$, $\probaa{\sigma}{\theta}{v}(\reach(\target) \mid \Gamma_w) \leq \val(w)$.
 %	Thus, if $\probaa{\sigma}{\theta}{v}(\reach(\target)) > 0$, $\val(v) \le \probaa{\sigma}{\theta}{v}(\reach(\target)) \le \max \{\val(w) \mid w\in Y \setminus Z\}$, in %contradiction with the definition of $Z$. Proposition~\ref{prop:rec} follows.
%je ne comprends pas
%pas besoin de preuve par l'absurde en fait
Hence $\probaa{\sigma}{\theta}{v}(\reach(\target)) \le \max (0,\max_{w\in Y\bh Z}\val(w))$
and since this holds for every $\sigma$, $\val(v)\leq \max (0,\max_{w\in Y\bh Z} \val(w))$.
By definition of $Z$ this implies $\val(v)=0$.
\end{proof}

\begin{prop}
\label{prop:valphi}
	Let $\vv$ be a live permutation such that:
\begin{equation*}%\label{eq:hypo}
			\val(\vv_1) \le \val(\vv_2) \le \ldots \le \val(\vv_k) \label{eq:ex0} \enspace .
\end{equation*}
Then $\vv$ is self-consistent.
\end{prop}

Note that under the same hypotheses, Lemma~\ref{lem:correctness} imply that $\vv$-strategies are optimal.
\begin{proof}
	We first show that:
	\begin{equation}
	 	\forall v \in \vertices, \forall 1 \le i \le k, \left(\vv(v) = i\right) \Rightarrow \left(\val(v) = \val(\vv_i)\right) \label{eq:ex1} \enspace .
	\end{equation}
	Consider the strategy $\sigma^*$, which mimics $\sigmav$ until the first time the pebble reaches a random vertex and then switches definitively to an optimal
	strategy. By definition of $\sigmav$, the first random vertex belongs to $\{\vv_i, \ldots, \vv_k, \target\}$, so $\sigma^*$ ensures that a pebble starting in $q$ reaches $\target$ with probability at least $\min \{\val(\vv_i),\ldots,\val(\vv_k),\val(\target)\} = \val(\vv_i)$. A similar strategy $\tau^*$ for $\m$ ensures that this probability is at most $\val(\vv_i)$. So $\val(v) = \val(\vv_i)$, and (\ref{eq:ex1}) follows.
	
	Now we prove that $\val$ and $\phi_\vv$ coincide.
	According to~\eqref{eq:ex1} and by definition of permutation strategies,
		\begin{align*}
			&&&\forall v \in \vM, & \val(v) &= \val(\sigmav(v)) \enspace
			%;\\
			%why ; not , like everybody?
			,\\
			&&&\forall v \in \vm, & \val(v) &= \val(\tauv(v)) \enspace
			%;\\
			%same
			,\\
			&&&\forall v \in \vr, & \val(v) &= \sum_{w\in\vertices} \tp{w}{v}\cdot\val(w)\enspace.
		\end{align*}
	So, if $\M$ and $\m$ play according to their $\vv$-strategies, the sequence $\val(V_i)_{i\in\NN}$ is a martingale:
		\begin{eqnarray}
			\esper{\sigmav}{\tauv}{v}{\val(V_{i+1})~\mid V_0\dots V_i} & = & \val(V_i)\enspace. \label{eq:ex2}
		\end{eqnarray}
	%We can then prove
	Consequenly, for every vertex $v$, $\phi_\vv(v)=\val(v)$:
		\begin{eqnarray}
		\phi_\vv(v) = \proba{\sigmav}{\tauv}{v}{\reach(\target)}	& = &	\esper{\sigmav}{\tauv}{v}{\lim_{i\in\NN} \val(V_i)}\label{eq:ex3}\\
								& = &	\lim_{i\in\NN} \esper{\sigmav}{\tauv}{v}{\val(V_i)}\label{eq:ex4}\\
								& = &	\esper{\sigmav}{\tau}{v}{\val(V_0)} = \val(v)\label{eq:ex5}\enspace,
		\end{eqnarray}
	where (\ref{eq:ex3}) comes from Proposition~\ref{prop:live}, (\ref{eq:ex4}) is a property of expectations, and (\ref{eq:ex5}) comes from (\ref{eq:ex2}).
	%(\ref{eq:ex0})
	%je n'ai pas compris la ref a eq:ex0	
	Since $\val$ and $\phi_\vv$ coincide, the hypothesis yields the self-consistency of $\vv$. This completes the proof of Proposition~\ref{prop:valphi}.
\end{proof}

\begin{lem}
\label{lem:existence}
	There exists a live and self-consistent permutation.
\end{lem}

\begin{proof}
	We use iteratively Proposition~\ref{prop:rec} in order to build a permutation $\vv$ such that, for every $k \ge i \ge 1$,
	\begin{enumerate}[$\bullet$]
		\item	$\val(\vv_i) = \max \left\{\val(v) \mid v \in \vertices \setminus \detatt{\vv_{i+1},\vv_{i+2},\ldots,\vv_k}\right\}$;
		\item	$\exists w \in \detatt{\vv_{i+1},\vv_{i+2},\ldots,\vv_k}, \tp{w}{\vv_i} > 0$.
	\end{enumerate}
	By construction $\vv$ is live and $\val(\vv_1) \le \val(\vv_2) \le \ldots \le \val(\vv_k)$. Proposition~\ref{prop:valphi} yields the self-consistency of $\vv$, and Lemma~\ref{lem:existence} follows.
\end{proof}

\section{The permutation-improvement algorithm}
\label{sec:heuristics}
A drawback of the permutation-enumeration algorithm is that it considers each and every possible permutation of the random vertices,
so $|\vr|!$ is a lower bound for the worst-case complexity of this algorithm. 
%il faut etre plks clair et se mettre plu sa la place d'un lecteur ovice
%hence there is no hope for the permutation-enumeration algorithm to perform well in practice.
Strategy-improvement algorithms avoid such enumerations,
instead these algorithms proceed by successive improvements of a strategy:
information about sub-optimality of a strategy
is used to determine a ``better'' strategy,
which ensures convergence to an optimal strategy.
%
%until an optimal strategy is discovered.
%becomes optimal.
%they gather information from
%incorrect strategies
%c'est quoi une strat correcte?
%non-optimal strategies in order to determine a ``better'' strategy in the next round.
In this section, we emulate this idea with a permutation-improvement algorithm.

\subsection{A natural but incorrect improvement policy}
\label{sec:naive}
\begin{figure}[ht]
	\begin{center}
\includegraphics{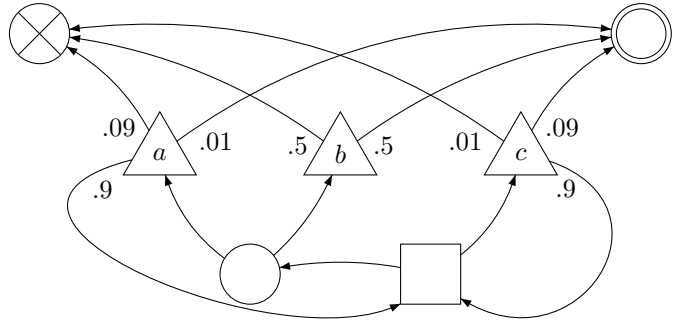}
%	\unitlength=.8mm
%		\begin{picture}(120,50)(0,5)
%		%\put(0,0){\framebox(120,60){}}
%			\gasset{polyangle=90}
%			\node(zero)(10,50){}
%			\drawline[AHnb=0](6.5,46.5)(13.5,53.5)
%			\drawline[AHnb=0](6.5,53.5)(13.5,46.5)
%		
%			\node[Nmarks=r](target)(110,50){}
%		
%			\rpnode(a)(30,30)(3,7){$a$}
%			\rpnode(b)(60,30)(3,7){$b$}
%			\rpnode(c)(90,30)(3,7){$c$}
%		
%			\node(eve)(45,10){}
%			\node[Nmr=0](adam)(75,10){}
%		
%			\drawedge[ELside=l,ELpos=25,ELdist=1,curvedepth=-3](a,zero){$.09$}
%			\drawedge[ELside=r,ELpos=10,ELdist=1,curvedepth=9](a,target){$.01$}
%			\drawbpedge[ELside=l,ELpos=10,ELdist=1](a,190,50,adam,245,24){$.9$}
%			
%			\drawedge[ELside=r,ELpos=12,ELdist=1,curvedepth=4](b,target){$.5$}
%			\drawedge[ELside=l,ELpos=12,ELdist=1,curvedepth=-4](b,zero){$.5$}
%		
%			\drawedge[ELside=r,ELpos=25,ELdist=1,curvedepth=3](c,target){$.09$}
%			\drawedge[ELside=l,ELpos=10,ELdist=1,curvedepth=-9](c,zero){$.01$}
%			\drawbpedge[ELside=r,ELpos=12,ELdist=1,curvedepth=5](c,-10,32,adam,-45,32){$.9$}
%		
%			\drawedge[ELside=l,ELpos=25,ELdist=1,curvedepth=3](eve,a){}
%			\drawedge[ELside=l,ELpos=25,ELdist=1,curvedepth=-2](eve,b){}
%		
%			\drawedge[ELside=l,ELpos=25,ELdist=1,curvedepth=-2](adam,eve){}
%			\drawedge[ELside=l,ELpos=25,ELdist=1,curvedepth=-3](adam,c){}
%	
%		\end{picture}
	\end{center}
	%attention les labels a l'interieur des caption
	\caption{\label{fig:naive} A counter-example for the na\"ive improvement algorithm.}
\end{figure}

%je rajoute une petite phrase d'intro pour huiler la lecture diagonale
Starting from an initial permutation $\vv$,
we would like to improve $\vv$ again and again
until the permutation strategies $\sigmav$ and $\tauv$ are optimal.
To test optimality
% with a permutation $\vv$,
we check that $\vv$
% $\vv$ 
is live and self-consistent (see Lemma~\ref{lem:correctness}).
When $\vv$ is live but \emph{not} self-consistent we compute a new permutation $\vvv$ which is live and``better'' than $\vv$.
A natural improvement policy consists in choosing $\vvv$ consistent with the $\vv$-values
i.e. $\vvv$ refines the pre-order induced by $\phi_\vv$.
Unfortunately this is too na\"ive: the corresponding algorithm does not always terminate\footnote{Actually, the na\"ive algorithm terminates (and is correct) in the special case of one-player games~\cite{Hor08-Thesis}.}, a counter-example is given by Figure~\ref{fig:naive}.

If we start with the permutation $\vv = acb$, the $\vv$-strategies are as follows: in $\fullmoon$ $\M$ goes to $b$ and in $\Box$ $\m$ goes to $c$. Hence, the $\vv$-values of vertices $a$, $c$, and $b$ are respectively $.82$, $.9$, and $.5$, so $\vv$ is not self-consistent. The next permutation is $\vvv = bac$, and the following $\vvv$-strategies ensue: in $\fullmoon$ $\M$ goes to $a$ and in $\Box$ $\m$ goes to $\fullmoon$. The $\vvv$-values of vertices $b$, $a$, and $c$ are respectively $.5$, $.1$, and $.18$, so $\vvv$ is not self-consistent either.
Moreover, the next permutation is $\vv = acb$, so the na\"ive algorithm oscillates endlessly between $\vv$ and $\vvv$, never reaching the correct permutation $abc$.

%\subsection{Mixed policy}
\subsection{A correct improvement policy}
\label{sec:improvement}
	The correct permutation-improvement policy is a bit more complex: given a live but not self-consistent permutation $\vv$, we choose a permutation $\vvv$ which is live and self-consistent in the one-player game $\game[\sigmav]$, where vertices of player $\M$ have only one outgoing edge: the edge consistent with the positional
	$\vv$-strategy $\sigmav$.
This improvement policy guarantees that the value of $\sigmavv$ is greater than the value of $\sigmav$ (see Lemma~\ref{lem:progress})
and is implemented by the following algorithm.

\begin{algorithm}[ht]
	\KwIn{A normalised simple stochastic game $G=(\vertices,\vM,\vm,\vr,\edges,\tpseul,\target,\sink)$.}
	\KwOut{Optimal strategies for $\M$ and $\m$.}
		Pick a live permutation $\vv$\;
		\REPEAT{}{
			\eIf{$\vv$ is self-consistent in $G$}{
				\Return{$\sigmav$ and $\tauv$}\;
			}{
				replace $\vv$ with a live and self-consistent permutation in $G[\sigmav]$ \label{line:improve}\;
			}
		}
	\caption{The permutation-improvement algorithm.}
	\label{alg:improvement}
	\end{algorithm}

The computation of a live and self-consistent permutation in $G[\sigmav]$ in line~\ref{line:improve}
relies on the computation of values of the one-player game $G[\sigmav]$.
Details are given in the proof of the following theorem.

	\begin{thm}
	\label{thm:improve}
	The permutation-improvement algorithm terminates and returns optimal strategies for $\M$ and $\m$ in at most $|\vr|!$ improvement steps. Furthermore, each improvement step can be carried out in polynomial time.
	\end{thm}

\begin{proof}
	%Proposition~\ref{prop:subgame} shows the soundness of line~\ref{line:improve}.
	%soundness je ne comprends pas cette histoire de soundness
	%comment le lecteur est-il censé deviner comment calculer la nouvelle permutation?
		According to Lemma~\ref{lem:doubleliveness} the algorithm returns a permutation
		which is both live and self-consistent in $G$ hence according to Lemma~\ref{lem:correctness}
		the corresponding permutation strategies are optimal in $G$ which proves correctness of the algorithm.

Termination and the maximal number of iterations follows from Lemma~\ref{lem:progress},
which proves that sucessive strategies $\sigmav$ have better and better values.

The computation of a live and self-consistent permutation in $G[\sigmav]$ in line~\ref{line:improve}
is achieved in polynomial time in the following way.
% of the algorithm can be achieved in polynomial time.
First, compute the values of the one-player game $G[\sigmav]$
using linear programming~\cite{HK79-MS,Con93-DMTCS}.
Second, build in linear time a live permutation $\vvv$
consistent with these values like in the proof of Lemma~\ref{lem:existence}.
The permutation $\vvv$ is such that $\val_{\sigmav}(\vvv_1)\leq \val{\sigmav}(\vvv_2)\leq \ldots \leq \val{\sigmav}(\vvv_k)$,
where $\val_{\sigmav}$ denotes the values in the game $G[\sigmav]$.
According to Proposition~\ref{prop:subgame} the game $G[\sigmav]$ is normalised hence Proposition~\ref{prop:valphi}
guarantees that $\vvv$ is consistent in $G[\sigmav]$.
\end{proof}

Let us compare briefly the permutation-enumeration and the permutation-improvement algorithms.
Each improvement step of the permutation-improvement
algorithm requires the computation of values
of a one-player SSG,
which can be performed using linear programming.
These values could be computed as well using a permutation-improvement policy or a strategy-improvement algorithm
in order to avoid linear programming altogether.
Either way, we have to forfeit one of the advantages of the permutation-enumeration algorithm: the computational simplicity of its inner loop.
On the other hand, we do not know any non-trivial lower bound on the number of loops in a run of the permutation-improvement algorithm: it may be polynomial.

%	Even before proving that line~\ref{line:improve} is sound, we have to show that $\vv$ remains invariably live in $G$ during a run.
	
\subsection{Soundness and correctness of the permutation-improvement algorithm}
The correctness proof is based on the following two results.

	\begin{lem}
	\label{lem:doubleliveness}
		Let $\sigma$ be a positional strategy for $\M$ and $\vv$ be a permutation. If $\vv$ is live in $G[\sigma]$ it is also live in $G$.
	\end{lem}

	\begin{proof}
		Let $W_\vv$ and $X_\vv$ denote the $\vv$-regions in $G$ and $G[\sigma]$, respectively. By definition, $\cup_{j>i} W_\vv[j]$ is the deterministic attractor of $\M$ to $\{\vv_i,\ldots,\vv_k,\target\}$ in $G$, while $\cup_{j>i} X_\vv[j]$ is the same attractor in $G[\sigma]$. As the moves of $\M$ are restricted in $G[\sigma]$, we get 
			\begin{equation}
				\forall 1\le i \le k, \bigcup_{j>i} X_\vv[j] \subseteq \bigcup_{j>i}W_\vv[j] \enspace . \label{eq:double1}
			\end{equation}
		Thus, the liveness of $\vv$ in $\game$ follows from its liveness in $\game[\sigma]$, and Lemma~\ref{lem:doubleliveness} ensues.
	\end{proof}

%	We can now show that line~\ref{line:improve} is sound. It follows ostensibly from Lemma~\ref{lem:existence}, but we have to show that $G[\sigmav]$ is normalised.

	\begin{prop}
	\label{prop:subgame}
		Let $\vv$ be a live permutation. Then $G[\sigmav]$ is normalised.
	\end{prop}
	\begin{proof}
		In the proof of Proposition~\ref{prop:live}, we have shown the existence of a positive real number $\alpha$ such that for any strategy $\tau$ for min and vertex $v \neq \sink$, $\probaa{\sigmav}{\tau}{v}(V_n = \target) \ge \alpha^k$ hence only $\sink$ has value $0$ in $G[\sigmav]$. Clearly only $\target$ has value $1$ in $G[\sigmav]$ hence Proposition~\ref{prop:subgame} follows.
	\end{proof}

\subsection{Termination of the permutation-improvement algorithm}

	The value of a strategy $\sigma$ is denoted $\val_\sigma$ and defined by:
	\[
	\forall v\in\vertices, \val_\sigma(v) =\inf_\tau \proba{\sigma}{\tau}{v}{\reach(\target)}\enspace.
	\]
	%la def de la valeur est bien ici, colle au premier endroit ou on l'utilise
	 %bof le changement val(sigma) vers val_sigma
	 %les doubles subscripts sont a eviter mais je change pas
	 For proving termination of the permutation-improvement algorithm we prove that successive strategies $\sigmav$
	chosen by the algorithm have greater and greater values.
 	\begin{lem}
 	\label{lem:progress}
		Let $\vv$ be a live permutation in $G$ and $\vvv$ be a live and self-consistent permutation in $G[\sigmav]$.
		Then for all $v \in \vertices$,
		\begin{equation}\label{eq:morre}
		 \val_{\sigmav}(v) \le \val_{\sigmavv}(v) 
		 \enspace.
		\end{equation}
		Moreover, if for all $v \in \vertices$, $\val_{\sigmavv}(v) = \val_{\sigmav}(v)$ then $\vvv$ is self-consistent in $G$.
	\end{lem}

	\begin{proof}
	A key remark in the proof of Lemma~\ref{lem:progress} is that:
	\begin{equation}
	\val_{\sigmav}(\vvv_1)\le\val_{\sigmav}(\vvv_2)\le\ldots\le\val_{\sigmav}(\vvv_k)\enspace \label{eq:coherence}.
	\end{equation}
	Let $\psi_{\vv,\vvv}$ be the $\vvv$-values in $G[\sigmav]$. The self-consistency of $\vvv$ in $G[\sigmav]$ is:
	\begin{equation*}
	\psi_{\vv,\vvv}(\vvv_1)\le \psi_{\vv,\vvv}(\vvv_2) \le \ldots \le \psi_{\vv,\vvv}(\vvv_k) \enspace .
	\end{equation*}
	Lemma~\ref{lem:correctness} applied to $G[\sigmav]$ implies that the $\vvv$-strategy of player $\m$
	in $G[\sigmav]$ is optimal in $G[\sigmav]$ hence $\psi_{\vv,\vvv} = \val_{\sigmav}$ and (\ref{eq:coherence}) follows.

%Tu ne guides pas le lecteur enfin tant pis
	Consider now the sequence $(\sigma^n)_{n\in\NN}$, where $\sigma^n$ is the strategy where $\M$ plays according to $\sigmavv$
	until the pebble has visited $n$ random vertices, and plays according to $\sigmav$ afterwards. In particular $\sigma^0 = \sigmav$.
	We show that for every vertex $v$ the sequence $(\val_{\sigma^n}(v))_{n\in\NN}$
	% is waxing
	% d'apres mon dico ca ne se dit que pour la lune
	is non-decreasing and that its limit is less than $\val_{\sigmavv}(v)$.
	%je patche
	Since $\sigma^0 = \sigmav$ this will prove Lemma~\eqref{eq:morre}.

	We first show by induction that for any integer $n$,
	\begin{equation}
		\forall v\in\vertices, \val_{\sigma^{n+1}}(v) \ge \val_{\sigma^n}(v) \label{eq:ii}\enspace .
	\end{equation}
	\textit{Basis ($n=0$):} We have to prove that values of $\sigma^1$ are greater than values of $\sigma_\vv$.
	% want to show:
	%\begin{equation}
	%	\forall v\in\vertices, \val_{\sigma^{1}}(v) \ge \val_{\sigmav}(v) \label{eq:ii0}\enspace .
	%\end{equation}
	Let $v$ be a vertex, $i$ be the index of the $\vvv$-region of $v$ in $\game$, and $j$ be the index of the $\vvv$-region of $v$ in $\game[\sigmav]$. As the moves of $\M$ are restricted in $\game[\sigmav]$
	%,
	 the definition of the $\vvv$-regions gives $i \ge j$
	 %,
	  and (\ref{eq:coherence}) yields:
	\begin{equation}
		\val_{\sigmav}(\vvv_i) \geq \val_{\sigmav}(\vvv_j)\enspace. \label{eq:ii1}
	\end{equation}
	If $\M$ plays with $\sigma^1$, the definition of $\sigmavv$ ensures that the first random vertex belongs to $\{\vvv_i,\vvv_{i+1},\ldots,\vvv_k,\target\}$, so
	$\val_{\sigma^1}(v) \geq \min\{\val_{\sigmav}(\vvv_i),\val_{\sigmav}(\vvv_{i+1}),\ldots,\val_{\sigmav}(\vvv_k),1\}$
	and~\eqref{eq:coherence} yields:
	\begin{equation}
	 \val_{\sigma^1}(v)\geq \val_{\sigmav}(\vvv_i)\enspace. \label{eq:ii2}
	\end{equation}
	On the other hand we prove:
	\begin{equation}
		\val_{\sigmav}(v) = \val_{\sigmav}(\vvv_j)\enspace. \label{eq:ii3}
	\end{equation}
	Let $\psi_{\vv,\vvv}$ denote $\vvv$-values in $G[\sigmavv]$.
	We have already proved above that $\psi_{\vv,\vvv}$ is equal to $\val_{\sigmav}$.
	By definition of $j$, $\vvv_j$ is the first random vertex in a play in $G[\sigmav]$ starting from $v$ and consistent with
	a $\vvv$-strategy for $\m$ in $G[\sigmav]$ hence $\psi_{\vv,\vvv}(v)=\psi_{\vv,\vvv}(\vvv_j)$ which yields~\eqref{eq:ii3}.
	
	It follows from (\ref{eq:ii1}), (\ref{eq:ii2}), and (\ref{eq:ii3}) that (\ref{eq:ii}) holds for $n = 0$.

	\noindent\textit{Inductive step ($n \Rightarrow n+1$):} The strategies $\sigma^{n+2}$ and $\sigma^{n+1}$ coincides with $\sigmavv$ until the first visit to a random vertex. Then $\sigma^{n+2}$ switches to $\sigma^{n+1}$ while $\sigma^{n+1}$ switches to $\sigma^{n}$. By induction hypothesis, $\val_{\sigma^{n+1}} \ge \val_{\sigma^{n}}$, so $\val_{\sigma^{n+2}} \ge \val_{\sigma^{n+1}}$ and (\ref{eq:ii}) holds for $n+1$.

	Now we show that for every $v$, $\lim_{n\in\NN}\val_{\sigma^n}(v)\leq \val_{\sigma_\vvv}(v)$. Let $\tau$ be a strategy for $\m$. We have:
	\begin{eqnarray}
	\proba{\sigmavv}{\tau}{v}{\reach(\target)} 	& = & \proba{\sigmavv}{\tau}{v}{\neg\reach(\sink)}\enspace, \label{eq:pro1}\\
						\notag	& = & \lim_n \proba{\sigmavv}{\tau}{v}{\forall m\leq n, V_m \neq \sink} \enspace, \\
							& = & \lim_n \proba{\sigma^{n}}{\tau}{v}{\forall m\leq n, V_m \neq \sink} \enspace, \label{eq:pro3}\\
							& \geq & \lim_n \proba{\sigma^{n}}{\tau}{v}{\reach{\target}} \geq \lim_n \val_{\sigma^{n}}(v)\enspace,\label{eq:pro4}
	\end{eqnarray}
	where (\ref{eq:pro1}) follows from Proposition~\ref{prop:live},  (\ref{eq:pro3}) holds because $\sigmavv$ coincides with $\sigma^n$ for at least $n$ steps, and (\ref{eq:pro4}) by event inclusion and by definition of the value. This holds for every strategy $\tau$ hence $\val_{\sigmavv}(v)\geq \lim_n \val_{\sigma^n}(v)$.
	
	Altogether, $\val_{\sigmav}(v) = \val_{\sigma^0}(v) \le \val_{\sigma^1}(v)\le\cdots\le \lim_n \val_{\sigma^n}(v) \le \val_{\sigmavv}(v)$ hence (\ref{eq:morre}),
	which achieves to prove the first part of the lemma.
	
	Let us suppose now that $\val_{\sigmav} = \val_{\sigmavv}$. Equation (\ref{eq:coherence}) yields:
	\begin{equation}
		\val_{\sigmavv}(\vvv_1)\le\val_{\sigmavv}(\vvv_2)\le\ldots\le\val_{\sigmavv}(\vvv_k)\enspace .\label{eq:selfcoherence}
	\end{equation}
	We can thus apply Proposition~\ref{prop:valphi} to $\vvv$ in 
	 $\game[\sigmavv]$ which yields the self-consistency of $\vvv$ in $\game[\sigmavv]$.
	 By definition of $\vvv$-zones, they coincide in $\game$ and $\game[\sigma_\vvv]$ hence the $\vvv$-values are equal in $\game$ and $\game[\sigmavv]$
	 and $\vvv$ is also self-consistent in $\game$. 
	\end{proof}

\section*{Conclusion}
	We have presented two algorithms computing optimal strategies in simple stochastic games: the permutation-enumeration and the permutation-improvement algorithms. Both of them rely on the existence of optimal permutation strategies. The permutation-enumeration algorithm simply tests every permutation until it finds a live and self-consistent one. The permutation-improvement algorithm uses a smarter policy in order to choose a ``better'' permutation in the next round, \emph{\`a la} Hoffman-Karp.
	
	The permutation-enumeration algorithm has exponential worst-case complexity but it is a witness that solving SSGs is fixed-parameter tractable when the parameter is the number of random vertices. The nominal complexity of the permutation-improvement algorithm is a bit higher but we do not know any non-trivial lower
bound on the number of improvement steps: the permutation-improvement algorithm may actually run in polynomial time.

Whether simple stochastic games are solvable in polynomial time remains a challenging open question.

\medskip

\noindent\textbf{Acknowledgements} We would like to thank Marcin
Jurdzi\'nski for some fruitful discussions, the anonymous reviewers
for several useful suggestions and Julien Cristau for his invaluable
comments during the writing of the final version.

\end{document}